\documentclass[amsmath,amsfonts,amssymb]{revtex4} 

\usepackage{amsmath}
\usepackage{amssymb}
\usepackage{amsthm}
\usepackage{mathrsfs}

\usepackage{ifthen}

\usepackage{natbib}

\usepackage[usenames,dvipsnames]{color}
\usepackage[a4paper, colorlinks=true]{hyperref}

\hypersetup{
   pdftitle = {Spectral Analysis of Radial Dirac Operators in the Kerr-Newman Metric and its Applications to Time-periodic Solutions},
   pdfauthor = {M. Winklmeier, O. Yamada},
   anchorcolor = red,
   linkcolor = Maroon,
   citecolor = OliveGreen,
   urlcolor = Sepia,
} 

\definecolor{refkey}{named}{CadetBlue}
\definecolor{labelkey}{named}{Mahogany}

\renewcommand{\i}{i}
\renewcommand{\cal}{\mathcal}
\newcommand{\rd}{d}
\newcommand{\Ltwo}{\mathscr L^2}
\newcommand{\Cinfty}{\mathscr C_0^\infty}

\DeclareMathOperator{\sign}{sign}

\newcommand{\hs}[1]{%
\ifthenelse{\equal{#1}{1}}{\mathcal H_{r,\theta,\varphi}}{
\ifthenelse{\equal{#1}{2}}{\mathcal H_{r,\theta}}{
\ifthenelse{\equal{#1}{3}}{\mathcal H_{x}}{
   \ifthenelse{\equal{#1}{4}}{\mathcal H_{\varphi}^{r,\theta}}{\mathcal H} } } }
}

\newcommand{\pdiff}[2][]{\frac{\partial #1}{\partial #2}}
\newcommand{\diff}[2][]{\frac{d #1}{d #2}}

\newtheorem{theorem}{Theorem}[section]
\newtheorem{lemma}[theorem]{Lemma}
\newtheorem{proposition}[theorem]{Proposition}

\newtheoremstyle{myRemark}{8.0pt plus 2.0pt minus 4.0pt}{8.0pt plus 2.0pt minus 4.0pt}{}{}%
   {\bfseries}
   {.}
   {5pt plus 1.0pt minus 1.0pt}
   {}

\theoremstyle{myRemark}
\newtheorem{remark}[theorem]{Remark}
\newtheorem{definition}[theorem]{Definition}

\begin{document}

\title{Spectral Analysis of Radial Dirac Operators in the Kerr-Newman Metric and its Applications to Time-periodic Solutions}

\date{\today} 

\author{Monika Winklmeier}
\affiliation{Fachbereich 3 -- Mathematik, 
   Universit\"{a}t Bremen, D-28359 Bremen, Germany}
\email{winklmeier@math.uni-bremen.de}

\author{Osanobu Yamada}
\affiliation{Department of Mathematical Sciences, 
   Ritsumeikan University, Kusatsu, Shiga 525--8577, Japan}
\email{yamadaos@se.ritsumei.ac.jp}

\begin{abstract}
   We investigate the existence of time-periodic solutions of the Dirac equation in the Kerr-Newman background metric.
   To this end, the solutions are expanded in a Fourier series with respect to the time variable $t$ and 
   the Chandrasekhar separation ansatz is applied so that the question of existence of a time-periodic solution is 
   reduced to the solvability of a certain coupled system of ordinary differential equations.
   First, we prove the already known result that there are no time-periodic solutions in the non-extreme case.
   Then it is shown that in the extreme case for fixed black hole data there is a sequence of particle masses 
   $(m_N)_{N\in\mathbb N}$ for which a time-periodic solution of the Dirac equation does exist. 
   The period of the solution depends only on the data of the  black hole described by the Kerr-Newman metric.
\end{abstract}



\maketitle

\section{Introduction}	
In this paper we consider the system consisting of a Kerr-Newman black hole and an electron. 
The Kerr-Newman metric describing the black hole is the most general electrovac solution
of Einstein's field equations.
It describes a rotating, electrically charged, massive black hole, see, e.g., 
\citep{wald} and \citep{FN98}.
We are interested in the stability of a system consisting of such a black hole 
and an electron. 
To this end, we have to consider the Dirac equation for the electron in the Kerr-Newman metric, 
see equations \eqref{time-dependent equation} and \eqref{original R}--\eqref{Delta}.
The Dirac equation is a complicated system of partial differential equations in all four spacetime variables.
We call the system consisting of the black hole and the electron {\em stable} if there exists a nontrivial time-periodic solution of the Dirac equation which can be interpreted as the wave function of the electron.

Since the black hole is rotating, the background metric is only axisymmetric, whereas in the non-rotating cases 
(the Schwarzschild and the Reissner-Nordstr{\o}m geometries) the background metric is spherically symmetric.
This loss of symmetry leads to a complicated coupling of the angular and the radial coordinates in the Dirac equation.
\citet{chandrasekhar} showed that, in spite of this complicated coupling, the Dirac equation can be separated 
into a system of ordinary differential equations, the so-called angular equation \eqref{eigenequation theta} and the 
radial equation \eqref{eigenequation r}. 
These differential equations have realisations as eigenvalue equations in appropriate Hilbert spaces. 
For the radial equation, $\omega = \nu\omega_0$ plays the role of the eigenvalue parameter. The eigenvalue $\omega$ has the physical interpretation as the energy of the electron.
It should be emphasised that, due to the lack of spherical symmetry of the spacetime, the eigenvalues of the radial and the angular equation are intertwined in a highly complex way.
Hence to show the stability of the system under consideration, it does not suffice to find eigenvalues of the radial 
and the angular equation separately, but we need to show that the eigenvalues are compatible with each other.

In section~\ref{sec:separation} we present the separation ansatz for time-periodic solutions of the Dirac equation 
in the Kerr-Newman background metric due to Chandrasekhar with mathematical rigour and we derive the radial and 
angular equation. 
In the next section we consider the radial equation.
It has been shown by \citet{BM} that the essential spectrum of the radial operator covers the real axis. 
\citet{FKSY} show the nonexistence of time-periodic solutions of the Dirac equation in the non-extreme Kerr-Newman geometry.
\citet{S} investigates the Dirac equations in the extreme Kerr-Newman geometry with the help of special functions and shows the existence of bounded states in the extreme Kerr case, which implies the existence theorem of time-periodic solutions. 
Our aim is to study the difference between the extreme and non-extreme case from the viewpoint of spectral analysis of the radial Dirac operators. 
We give also the existence theorem of time-periodic solutions in the extreme Kerr-Newman case.
%
%
In Theorem~\ref{theorem:nonexistence} we show that in the non-extreme Kerr-Newman metric the Dirac equation has
no time-periodic solution that has an interpretation as a particle wave function.
The case of the extreme Kerr-Newman metric is investigated in section~\ref{sec:ExtremeCase}.
It turns out that in this case there may be an eigenvalue of the radial equation.
\citet{S} proved a sufficient condition for the existence of a time-periodic solution of the Dirac equation.
However, it is not easy to verify that for given particle data there are black hole parameters such that this condition can be satisfied.
This problem is discussed in section~\ref{sec:energy_eigenvalues}.
We show that for fixed data of an extreme Kerr-Newman black hole 
there is a sequence of particle masses such that the system consisting of the quantum particle and the black hole 
permits time-periodic solutions.


\section{Separation of the Dirac equation in the Kerr-Newman background metric}	
\label{sec:separation}
We consider the Dirac equation (see, e.g., \citet{Page}, \citet{chandrasekhar})
\begin{equation}\label{time-dependent equation}
   (\widehat{{\cal R}}+\widehat{{\cal A}})\widehat{\Psi}\ =\ 0 
\end{equation}
for a spin-$\frac{1}{2}$ particle with the mass $m\ge0$ and the charge $e$ in the Kerr-Newman geometry,
where 
\allowdisplaybreaks{
\begin{eqnarray}
   \widehat{{\cal R}} &:=& 
      \left(\begin{array}{cccc} 
      imr &    0 & \sqrt{\Delta} \,{\cal D}_+ & 0                        
      \\
      0   & -imr & 0                        & \sqrt{\Delta}\,{\cal D}_- 
      \\
      \sqrt{\Delta}\,{\cal D}_- & 0   & -imr & 0 
      \\
      0 & \sqrt{\Delta}\,{\cal D}_+   & 0  & imr  
      \end{array}\right), \label{original R} 
   \\[2ex]
   \widehat{{\cal A}} &:=& 
      \left(\begin{array}{cccc} 
      -am\cos{\theta} &    0 & 0 & {\cal L}_+  
      \\
      0   & am\cos{\theta} & -{\cal L}_- & 0 
      \\
      0 & {\cal L}_+ & -am\cos{\theta} & 0 
      \\
      -{\cal L}_- & 0 & 0 & am\cos{\theta} 
      \end{array}\right),  \label{original A}
   \\[2ex]
   {\cal D}_\pm &:=& 
      \pdiff{r}\mp \frac{1}{\Delta}\left[(r^2+a^2) 
      \pdiff{t} + a \pdiff{\varphi} -ieQr
      \right] , 
   \\[2ex]
   {\cal L}_\pm &:=& 
      \pdiff{\theta}+ \frac{\cot{\theta}}{2} \mp i 
      \left[a \sin{\theta} \pdiff{t}+\frac{1}{\sin{\theta}} 
      \pdiff{\varphi}\right], 
   \\[2ex]
   \Delta(r) &:=& r^2-2Mr+a^2+Q^2
   \label{Delta}
\end{eqnarray} 
}
and $\widehat\Psi$ is the wave function of the spin-$\frac{1}{2}$ particle under consideration.
If the so-called black hole condition 
\begin{equation}
   M^2-a^2 -Q^2 \ \ge\ 0
   \label{black hole condition}
\end{equation}
holds, then the Kerr-Newman geometry is interpreted as the spacetime geometry generated by a black hole  with the mass 
$M\ge 0$, the electric charge $Q$ and the angular moment $J$; 
if $M\neq 0$, then the so-called Kerr parameter $a=J/M$ is the angular momentum of the black hole per unit mass.
The black hole condition~\eqref{black hole condition} ensures that the function $\Delta$ can be written as the product
\begin{equation}
   \Delta(r)\ =\ (r-r_+)(r-r_-) \tag{\ref{Delta}$'$}
   \label{Delta'}
\end{equation}
with
\begin{equation}
   r_\pm\ =\ M\pm \sqrt{M^2-a^2-Q^2}.
\end{equation}
The special case $M^2-a^2-Q^2=0$, that is the case where  $r_+=r_- = M$,
is referred to as the extreme Kerr-Newman metric. 

Let us recall that the Kerr-Newman metric is the most general electrovac solution of Einstein's field 
equations~\citep{FN98}. 
Special cases contained in the Kerr-Newman geometry are the Kerr geometry (if $Q=0$), the Reissner-Nordstr{\o}m 
geometry (if $a=0$) and the Schwarzschild geometry (if $Q=0$ and $a=0$).

A solution $\widehat \Psi$ of~\eqref{time-dependent equation} for
$(r, \theta, \varphi, t)\in\widehat\Omega := (r_+,\infty)\times(0,\pi)\times(-\pi,\pi)\times(-\infty,\infty)$
such that for every fixed time $t$ the function $\widehat\Psi(\cdot, \cdot, \cdot, t)$ lies in a suitable $\Ltwo$-space $\hs{1}$ (see \eqref{hs1}) 
can be interpreted as the wave function of the electron. 
Hence the existence of such a $\widehat\Psi$ would imply that the system consisting of the black hole and the spin-$\frac{1}{2}$ particle in its exterior is stable.

\begin{remark}[Dirac equation in flat space time]
   In the case of flat spacetime, i.e. for $a=0$, $M=0$, $Q=0$, the Dirac equation given in  \eqref{time-dependent equation} is unitarily equivalent to the familiar Dirac equation 
   \begin{align*}
      \Bigl(-i \pdiff{t} -  i\, \vec\alpha\cdot \vec\nabla + \beta m \Bigr)\Psi = 0
   \end{align*}
   as given, for instance, in \citet{Davydov}.
   For the proof we refer to \citet{W}.
\end{remark}

\subsection{Time periodic solutions}
In this paper we consider time-periodic solutions $\widehat\Psi$, that is, solutions such that 
\[\widehat{\Psi}(r,\theta,\varphi,t)= \widehat\Psi(r,\theta,\varphi, t+\textstyle\frac{2\pi}{\omega_0}) \hspace{6ex} 
   ((r, \theta, \varphi, t)\in \widehat\Omega)
      \]
for some $\omega_0>0$. 
In this case, the solutions can be expanded in a Fourier series
\begin{align*}
   \widehat\Psi(r, \theta,\varphi,t)\ =\ 
   \sum\limits_{\nu\in\mathbb Z} \exp(-\i \nu\omega_0t)\, \Psi^\nu(r,\theta,\varphi).
\end{align*}
For physical reasons, each wave function $\Psi^\nu$ must be an element of the Hilbert space
\begin{align}\label{hs1}
   \hs{1} \ :=\
   \Ltwo\Bigl((r_+,\infty)\times (0,\pi) \times (-\pi,\pi);\ 
      \textstyle\frac{r^2+a^2}{\Delta(r)}\, \sin{\theta}\ \rd r\,\rd\theta\,\rd\varphi\Bigr)^4
\end{align}
with the inner product according to \citet{FKSY}
\begin{align}
   \label{inner product}
   (\Psi, \Phi)\ =\ \int_{r_+}^\infty\int_0^\pi\int_{-\pi}^\pi 
   \langle\Psi(r,\theta,\varphi)\,,\, \Phi(r,\theta,\varphi)\rangle_{\mathbb C^4}\, \frac{r^2+a^2}
   {\Delta(r)} \sin{\theta}\,dr\,d\theta\,d\varphi .
\end{align}
Here 
$\langle\, \cdot\,,\,\cdot\,\rangle_{\mathbb C^4}$ is the usual scalar product on $\mathbb C^4$.
In order to separate off also the $\varphi$-dependence of the solution $\Psi^\nu$ we use the complete system 
$\{\exp{(i\kappa\varphi)}\, :\, \kappa \in \mathbb Z +\frac{1}{2}\}$ in 
$\Ltwo((-\pi,\pi);\ d\varphi)$ to employ the ansatz
\begin{align*}
   \Psi^\nu(r,\theta,\varphi)\ =\ 
   \sum\limits_{\kappa\in\mathbb Z+ \frac{1}{2}}
   \exp{(-\i\kappa\varphi)}\Psi^{\nu,\kappa}(r,\theta).
\end{align*}
Thus, if 
\begin{align*}
   \widehat\Psi(r,\theta,\varphi,t) 
   = \sum\limits_{\nu\in\mathbb Z} \sum\limits_{\kappa\in\mathbb Z+\frac{1}{2}}
   \exp{(-\i\omega_0 \nu t)}\exp{(-\i\kappa\varphi)} \Psi^{\nu,\kappa}(r,\theta)
\end{align*}
satisfies~\eqref{time-dependent equation}, then each $\Psi^{\nu,\kappa}$ 
is a solution of 
\begin{equation} \label{k-dependent equation}
   ({\cal R}_{\nu,\kappa} + {\cal A}_{\nu,\kappa})\Psi^{\nu, \kappa} =0
   \qquad\text{on}\qquad (r_0,\infty)\times(0,\pi)
   ,
\end{equation}
where $\cal R_{\nu, \kappa}$ and $\cal A_{\nu,\kappa}$ are obtained from 
\eqref{original R} and \eqref{original A} by replacing $ {\cal D}_{\pm}$ and ${\cal L}_{\pm}$ by
\begin{eqnarray*}
   {\cal D}_{\pm, \nu, \kappa} &:=& 
   \pdiff{r}\pm \frac{\i}{\Delta}\left[\omega_0 \nu(r^2+a^2) +\kappa    a +eQr\right],  
   \\[2ex]
   {\cal L}_{\pm, \nu, \kappa} &:=& 
   \pdiff{\theta}+ \frac{\cot{\theta}}{2} \mp  \left[a\omega_0 \nu 
      \sin{\theta}+\frac{\kappa}{\sin{\theta}}\right],
\end{eqnarray*}
respectively (see the proof of Theorem~\ref{theorem:nonexistence}).

\subsection{Separation of the radial and the angular coordinate}
To study the equation (\ref{k-dependent equation}) we consider the formal differential expression
\[\mathfrak A_{\nu,\kappa} := 
   \left(\begin{array}{cc}
	 -am\cos{\theta}& {\cal L}_{-, \nu,\kappa} 
	 \\
	 - {\cal L}_{+,\nu,\kappa} & am \cos{\theta} \end{array}
\right)\]
for $\theta\in(0,\pi)$.
For any half integer $\kappa$ the differential expression $\mathfrak A_{\nu,\kappa}$ has a unique self-adjoint 
realisation $A_{\nu,\kappa}$ in the Hilbert space $\Ltwo((0,\pi);\ \sin{\theta}\,d\theta)^2$ which has purely
discrete spectrum 
$\sigma(A_{\nu,\kappa}) = \{\lambda_{\nu, \kappa,n}\, :\, n\in\mathbb Z\setminus\{0\}\}\subseteq\mathbb R$
where each eigenvalue is simple (see, e.g., \citet{BSW}, \citet{W}). 
Hence there is a complete set of orthonormal eigenfunctions of $A_{\nu,\kappa}$
\begin{equation}
   \label{angular eigenfunctions} 
   g^{\nu, \kappa,n}\ :=\
   \begin{pmatrix} g_1^{\nu,\kappa, n} \\[1ex] g_2^{\nu,\kappa,n} \end{pmatrix}\ 
   \in\ \Ltwo( (0,\pi);\ \sin\theta\,\rd\theta)^2 
   \hspace{5ex}  (n \in {\mathbb Z\setminus\{0\}})
\end{equation}
with eigenvalues $\lambda_{\nu,\kappa, n}$. 
The family~\eqref{angular eigenfunctions} allows us to make the ansatz
\begin{align*}
   \Psi^{\nu,\kappa}\, =\, \sum\limits_{n\in\mathbb Z\setminus\{0\}} \Psi^{\nu,\kappa,n}
\end{align*}
with
\[\Psi^{\nu,\kappa,n}(r,\theta)\ =\
   \begin{pmatrix}
      X_-^{\nu,\kappa,n}(r)\, g_2^{\nu,\kappa,n}(\theta) \\[1ex]
      X_+^{\nu,\kappa,n}(r)\, g_1^{\nu,\kappa,n}(\theta) \\[1ex] 
      X_+^{\nu,\kappa,n}(r)\, g_2^{\nu,\kappa,n}(\theta) \\[1ex] 
      X_-^{\nu,\kappa,n}(r)\, g_1^{\nu,\kappa,n}(\theta)  
   \end{pmatrix}
   \hspace{7ex}
   (r\in(r_+,\infty),\ \theta\in(0,\pi))
\]
which leads to a separation of the angular coordinate $\theta$ and the radial coordinate $r$ (cf. \citet{chandrasekhar}): 
the angular function $g^{\nu,\kappa,n}$ satisfies the angular equation
\begin{align}\label{eigenequation theta}
   A_{\nu,n} g^{\nu,\kappa,n}\, =\, \lambda_{\nu,\kappa,n} g^{\nu,\kappa,n}
\end{align}
with the integrability condition $g^{\nu,\kappa,n}\in\Ltwo( (0,\pi);\ \sin\theta\,\rd\theta)$ and 
the radial function $X^{\nu,\kappa, n} = {}^t(X^{\nu,\kappa,n}_+,\, X^{\nu,\kappa,n}_-)$ satisfies the radial equation
\begin{equation} \label{eigenequation r}
   \left(\begin{array}{cc} 
	 -imr-\lambda_{\nu,\kappa, n} & \sqrt{\Delta}\,{\cal D}_{-,\nu, \kappa} \\[1ex]
	 \sqrt{\Delta}\,{\cal D}_{+,\nu, \kappa} & imr -\lambda_{\nu,\kappa, n}  
   \end{array}  \right)
   \begin{pmatrix}
	 X^{\nu,\kappa,n}_+(r) \\[1ex] X^{\nu,\kappa,n}_-(r)   
   \end{pmatrix}
   \, =\, 0,
\end{equation}
with the integrability condition arising from the inner product on $\hs{1}$ 
\begin{equation} \label{b.c.} 
   \int_{r_+}^\infty \bigl(|X^{\nu,\kappa,n}_+(r)|^2+|X^{\nu,\kappa,n}_-(r)|^2\bigr)\, \frac{r^2+a^2}{\Delta(r)}\,\rd r\
   <\ \infty.
\end{equation}  

\subsection{The radial equation}
Let 
\begin{align}\label{eq:widetilde f}
   W := \frac{1}{\sqrt{2}}\begin{pmatrix} -\i & \i  \\ -1 & -1    \end{pmatrix},
   \hspace{3ex}
   V := \begin{pmatrix} 0 & - 1/\sqrt{\Delta} \\ 1/\sqrt{\Delta} & 0 \end{pmatrix},
   \hspace{3ex}
   \widetilde f^{\nu,\kappa,n}(r)\, :=\,
   \left(\begin{array}{c} 
	 \widetilde f^{\nu,\kappa,n}_1(r) \\ \widetilde f^{\nu,\kappa,n}_2(r)   
   \end{array}\right)
   \, :=\, W
   \left( \begin{array}{c} 
	 X^{\nu,\kappa,n}_+(r) \\ X^{\nu,\kappa,n}_-(r)
   \end{array}\right).
\end{align}
Then we obtain from (\ref{eigenequation r})
\begin{eqnarray}\label{transformed eigenequation r}
    0&=& VW\left(\begin{array}{cc} 
	 -imr-\lambda_{\nu,\kappa, n} & \sqrt{\Delta}\,{\cal D}_{-,\nu, \kappa} \\[1ex]
	 \sqrt{\Delta}\,{\cal D}_{+,\nu, \kappa} & imr -\lambda_{\nu,\kappa, n}  
   \end{array}  \right)W^{-1} \widetilde f^{\nu,\kappa,n}(r)
 \\[3ex]
 \nonumber
   &=& 
   \begin{pmatrix}
      {\displaystyle{\frac{mr}{\sqrt{\Delta}}-\frac{a\kappa+eQr+\omega_0 \nu(r^2+a^2)}{\Delta}} }
      & 
      {\displaystyle-\diff{r}+\frac{\lambda_{\nu,\kappa,n}}{\sqrt{\Delta}} }  
      \\[3ex]
      {\displaystyle{\diff{r}+\frac{\lambda_{\nu,\kappa,n}}{\sqrt{\Delta}} }} 
      & 
      \displaystyle{-\frac{mr}{\sqrt{\Delta}}-\frac{a\kappa+eQr+\omega_0 \nu(r^2+a^2)}{\Delta}}  
   \end{pmatrix}
       \widetilde f^{\nu,\kappa,n}(r).
\end{eqnarray}
for $r\in(r_+,\infty)$ (see \citet{W}). 

\begin{remark}
   If we take into account only terms of first order in $1/r$ for large $r$, then the radial equation becomes
   \begin{align*}
   \begin{pmatrix}
      \displaystyle{m - \frac{eQ}{r} + \omega_0 \nu}
      & 
      \displaystyle{-\diff{r} + \frac{\lambda_{\nu,n,\kappa}}{r} }  
      \\[3ex]
      \displaystyle{\diff{r}+\frac{\lambda_{\nu,n,\kappa}}{r} } 
      & 
      \displaystyle{-m - \frac{eQ}{r} + \omega_0\nu} 
   \end{pmatrix}
   \widetilde f(r) 
   =0,
   \end{align*}
   which is exactly the radial equation for the relativistic hydrogen atom, see, e.g., \citet{BD}.
\end{remark}
\begin{remark}\label{remark:a=0}
   It is important to note that in the case $a\neq0$ the eigenvalue $\lambda$ for the angular equation does depend on $\omega_0 \nu$.
   In the case $a=0$, the eigenvalues can be calculated explicitly; they are given by
   \begin{align}\label{spectrum for a=0}
         \textstyle
         \sigma(A_{\nu,\kappa})\,
         =\, \bigl\{
         \lambda_{\nu, \kappa,n} = \sign(n)  \bigl(|\kappa| -\frac{1}{2} + |n|\bigr)
         \, :\, n\in\mathbb Z\setminus\{0\}\bigr\}
         \, \subseteq\, \mathbb R.
   \end{align}
   In particular, the eigenvalues $\lambda_{\nu,\kappa,n}$ of the radial equation do not depend on the eigenvalues of the radial equation.
   Hence, in the case $a=0$, for a solution of the complete problem~\eqref{eigenequation r}--\eqref{eigenequation theta} first the angular 
   problem $(A_{\nu,\kappa}-\lambda_{\nu,\kappa})g=0$ is solved for the eigenvalues $\lambda_{\nu,\kappa,n}$, and then 
   the radial equation~\eqref{eigenequation r} can be attacked for fixed $\lambda_{\nu,\kappa,n}$.

\end{remark}
We introduce a new coordinate $x$ such that 
\begin{equation} \label{definition of x}
   \frac{\rd x}{\rd r}\ =\ \frac{r^2+a^2}{\Delta(r)} \hspace{5ex} (r>r_+),
\end{equation}
that is, 
\begin{equation}\label{precise x}
   x(r)\ =\
   \begin{cases}
      r+ \frac{r_+^2+a^2}{r_+-r_-}\log{(r-r_+)}-  \frac{r_+^2+a^2}{r_+-r_-}
      \log{(r-r_-)}   +x_0  
      \hspace{1ex}
      &(r_+\neq r_-),
      \\[2ex]
      r+ 2r_+\log{(r-r_+)}-\frac{r_+^2+a^2}{r-r_+} + x_0 
      &(r_+=r_-),
   \end{cases}
\end{equation}
where $x_0$ is a constant of integration that can be set $x_0=0$. 
The correspondence between $r>r_+$ and $x\in (-\infty,\infty)$ is a bijection. 
With the new coordinate $x$ equation~\eqref{transformed eigenequation r} becomes 
\begin{eqnarray} 
   H_{\nu,\kappa,n}f^{\nu,\kappa,n} &:=&
   \left(\begin{array}{cc} 
	 \displaystyle{\frac{mr\sqrt{\Delta}}{r^2+a^2}-\frac{a\kappa+eQr}{r^2+a^2}}      &
	 \displaystyle{ -\diff{x}+\frac{\lambda_{\nu,\kappa,n}\sqrt{\Delta}}
	    {r^2+a^2}}  
	 \\[2ex] 
	 \displaystyle{\diff{x}+ \frac{\lambda_{\nu,\kappa,n}\sqrt{\Delta}}
	    {r^2+a^2}} 
	 & 
	 \displaystyle{-\frac{mr\sqrt{\Delta}}{r^2+a^2}-\frac{a\kappa+eQr}{r^2+a^2}      } 
   \end{array}\right)f^{\nu,\kappa,n}\  \nonumber \\[3ex]
   &=& \omega_0 \nu f^{\nu,\kappa,n}, \label{definition of H}
\end{eqnarray}
where $r$ has to be understood as $r(x)$ and 
$f^{\nu,\kappa,n}(x)=\widetilde f^{\nu,\kappa,n}(r(x))$, $f^{\nu,\kappa,n}_j(x) = \widetilde f^{\nu,\kappa,n}_j(r(x))$ 
for all $x\in(-\infty, \infty)$ and $j\in\{1,\, 2\}$. 
In view of (\ref{eq:widetilde f}) and (\ref{definition of x}) the integrability condition (\ref{b.c.}) becomes 
\begin{align}\label{f int. cond}
   \int_{-\infty}^\infty \bigl\|\,f^{\nu,\kappa,n}(x)\,\bigr\|_{\mathbb C^2}^2\,dx\ =\
   \int_{-\infty}^\infty \bigl[\,\bigl|f^{\nu,\kappa,n}_1(x)\bigr|^2+\bigl|f^{\nu,\kappa,n}_2(x)\bigr|^2\,\bigr]\,dx\
   <\ \infty.
\end{align}
The operator $H_{\nu,\kappa,n}$ is formally symmetric in the Hilbert space $\hs{3}:=\Ltwo((-\infty,\infty);\, \rd x)^2$,
so it is natural to look for an operator theoretical realisation of $H_{\nu,\kappa,n}$ in the Hilbert space $\hs{3}$.
\bigskip
The purpose of this note is to investigate the spectral properties of the self-adjoint operator $H_{\nu,\kappa, n}$ 
in $\hs{3} = \Ltwo((-\infty,\infty);\ \rd x)^2$ and to study the non-existence of non-trivial solutions satisfying (\ref{k-dependent equation}). 
In section~\ref{sec:energy_eigenvalues} we will investigate the existence of so-called energy eigenvalues of $H_{\nu,\kappa,n}$ (cf.~\citet{S}).

\begin{definition}
   We call $\omega \in\mathbb R$ an {\em energy eigenvalue} of $H_{\nu, \kappa, n}$
   if there are $\lambda_{\nu,\kappa, n}$ such that $\omega$ is an eigenvalue of $H_{\nu,\kappa,n}$ and $\lambda_{\nu,\kappa, n}$ is an eigenvalue of $A_{\nu,\kappa}$, that is, if equations~\eqref{eigenequation theta} and \eqref{eigenequation r} can be solved simultaneously with functions satisfying the corresponding integrability conditions.
\end{definition}


\section{The operator $H$} 
\label{sec:operatorH}
In this and the following sections we consider the 
eigenvalue equation $(H_{\nu,\kappa,n}-\omega\nu_0) f^{\nu,\kappa,n} = 0$ from \eqref{definition of H} on 
the Hilbert space $\hs{3} = \Ltwo((-\infty,\infty);\ \rd x)^2$.

If there is no ambiguity, we omit the indices $\nu$, $n$ and $\kappa$ in the following for the sake of clarity; for instance, we write simply $H$ instead of $H_{\nu,\kappa,n}$, $\lambda$ instead of $\lambda_{\nu,\kappa,n}$ and $\omega$ instead of $\omega_0\nu$.
We decompose the operator $H$ into the sum
\[H = H_0 + V\] 
where
{\allowdisplaybreaks
\begin{eqnarray*}
   H_0 &=& 
   \left(\begin{array}{cc} 
	 0 &-\frac{\rd}{\rd x}
	 \\ \frac{\rd}{\rd x} & 0
   \end{array}\right), 
   \hspace{5ex} \mathcal D(H_0)\ =\ \Cinfty(-\infty, \infty)^2,
   \\[2ex]
   V(x)&=& 
   \left(\begin{array}{cc} 
	 A(x)& B(x) \\ B(x) & C(x) 
   \end{array}\right),
   \\[2ex]
   A(x) &=& 
   \frac{m\, r(x)\sqrt{\Delta(r(x))}}{r(x)^2+a^2} 
   - \frac{a\kappa+eQr(x)}{r(x)^2+a^2}, 
   \\[2ex]
   B(x) &=& 
   \lambda\, \frac{\sqrt{\Delta(r(x))}}{r(x)^2+a^2},
   \\ [2ex]
   C(x) &=& 
   - \frac{m\, r(x)\sqrt{\Delta(r(x))}}{r(x)^2+a^2} 
   - \frac{a\kappa+eQr(x)}{r(x)^2+a^2},
   \\[2ex]
   \Delta(x) &=&
   (r(x)-r_+)(r(x)-r_-).
\end{eqnarray*}
}
The operator $H_0$ is symmetric and has a unique self-adjoint extension on the space $\Ltwo((-\infty,\infty);\ dx)^2$, see, e.g., \citet[Theorem 6.8]{We87}.
Since $x \rightarrow - \infty$ is equivalent to $r(x) \rightarrow r_+$
and 
$x \rightarrow \infty$ is equivalent to $r(x) \rightarrow \infty$, we have
\begin{align*}
   \lim_{x\rightarrow -\infty} A(x) & = - \frac{a\kappa+eQr_+}{r_+^2+a^2} =:A_0,
   & \lim_{x\rightarrow +\infty} A(x) & = m, 
   \\[1ex]
   \lim_{x\rightarrow -\infty} C(x) &= A_0, 
   & \lim_{x\rightarrow +\infty} C(x) & = -m,
   \\[1ex]
   \lim_{x\rightarrow \pm \infty} B(x) &= 0
\end{align*}  
which implies that the functions $A(\cdot)$, $B(\cdot)$ and $C(\cdot)$ are bounded. 
Since we assume that the black hole condition~\eqref{black hole condition} holds, the multiplication operator $V$ is symmetric.
Therefore, $H=H_0+V$ has a unique self-adjoint extension which we again denote by $H$.

\par\medskip
In what follows, a prime $\ {}'\ $ always denotes differentiation with respect to $x$.

\begin{lemma}[Asymptotic behaviour of $V$ for $x\rightarrow-\infty$]\ 
   \label{lemma:A' for -infty}
   \\\vspace{-\topsep}\vspace{-\itemsep}
   \begin{enumerate}
      \item\label{item:asymptotics}
      For $x\rightarrow-\infty$ the functions 
      \[ A(x) -A_0, \ \  B(x), \ \  C(x)-A_0  \] 
      decay exponentially in the case $r_+\neq r_-$, and 
      they are of order $O(x^{-1})$ in the case $r_+=r_-$. More precisely, in the latter case we have
      \begin{alignat*}{3}
	 A(x)-A_0\ &=\ \frac{mM-\mu}{-x}+O\left(\frac{1}{x^2}\right)\hspace{4ex}
	 && {\rm as} \ \ x \rightarrow -\infty,
	 \\
	 A_0-C(x)\ &=\ \frac{mM+\mu}{-x}+O\left(\frac{1}{x^2}\right)
	 && {\rm as} \ \ x \rightarrow -\infty,
	 \\
	 B(x)\ &=\ \frac{\lambda}{\,x\,}+O\left(\frac{1}{x^2}\right)
	 && {\rm as} \ \ x \rightarrow -\infty,
      \end{alignat*}
      where
      \[\mu\ :=\ -\frac{2a\kappa M}{M^2+a^2} - eQ\frac{M^2-a^2}{M^2+a^2}. \]

      \item\label{item:integrability}
      The derivatives $A^\prime$, $B^\prime$ and $C^\prime$ are integrable with respect to $x$ on $(-\infty,0]$.  

      \item\label{item:A' for infty}
      $A'(x)=O(x^{-2})$, $B'(x)=O(x^{-2})$ and $C'(x)=O(x^{-2})$ hold as $x \rightarrow +\infty$. 

   \end{enumerate}
\end{lemma}

\begin{proof}
   We prove the assertions for the functions $B$ and $A$ only. The corresponding assertions for $C$ can 
   be obtained from those for $A$ by substituting $m$ by $-m$. 
   We remark that (\ref{precise x}) shows
   \begin{alignat}{3}
      \label{eq:r asymptotics nonextreme case}
      r(x)-r_+ &=\, O(\exp{(\alpha x)})\hspace{4ex}  
      &&\text{as}\hspace{2ex}  x \rightarrow -\infty
      &\hspace{4ex}&(r_+\neq r_-), \\ 
      \label{eq:r asymptotics extreme case}
      r(x)-r_+ &=\, O(x^{-1})
      &&\text{as}\hspace{2ex}  x \rightarrow -\infty
      &&(r_+ = r_-)
   \end{alignat}
   for a positive constant $\alpha$. 
   To keep notation simple, let us write $r$ instead of $r(x)$ in this proof.
   \begin{enumerate}
      \item 
      Note that 
      \begin{align}\label{eq:asymptotics}
	 \frac{1}{r^2+a^2}\ =\ \frac{1}{r_+^2+a^2} - \frac{ (r-r_+)(r+r_+) }{ (r_+^2 + a^2)(r^2 + a^2)}\
	 =\ \frac{1}{r_+^2+a^2} + O(r-r_+).
      \end{align}

      Hence it follows immediately that
      \begin{align*}
	 B(x)\ =\ \lambda \frac{\sqrt{\Delta(r)}}{r^2 + a^2}\ 
	 =\ \lambda \sqrt{\Delta(r)}\left( \frac{1}{r_+^2+a^2} + O(r-r_+) \right).
      \end{align*}
      If we recall that $\Delta(r(x))=(r(x)-r_+)(r(x)-r_-)$ and use the relations 
      \eqref{eq:r asymptotics nonextreme case} and \eqref{eq:r asymptotics extreme case}, we see that 
      the assertion for $B$ holds.

      Now we prove the assertions for $A$.
      Simple calculations show
      \begin{align*}
	 A(x) - A_0\  &=\
	 \frac{mr\sqrt{\Delta(r)}}{r^2 + a^2}
	 + \frac{ (r^2+a^2)(a\kappa + eQr_+) - (r_+^2+a^2)(a\kappa + eQr) }{(r^2+a^2)(r_+^2+a^2)}
	 \\[2ex]
	 &\begin{alignedat}{2}
	    & =\  \frac{mr\sqrt{\Delta(r)}}{r^2 + a^2}
	    + \frac{ (r-r_+)\left[ (r-r_+)(a\kappa + eQr_+) + 2a\kappa r_+ + eQ(r_+^2 - a^2) \right] }{(r^2+a^2)(r_+^2+a^2)}
	    \\[2ex]
	    & =\
	    \frac{mr_+\sqrt{\Delta(r)}}{r^2 + a^2}
	    + \frac{r-r_+}{r^2+a^2} \left[
	       m\sqrt{\Delta(r)}
	       + \frac{ (r-r_+)(a\kappa + eQr_+) + 2a\kappa r_+ + eQ(r_+^2 - a^2) }{r_+^2+a^2}
	    \right].
	 \end{alignedat}
      \end{align*}

   Using the relations \eqref{eq:asymptotics} and \eqref{eq:r asymptotics nonextreme case}, we see that the assertion holds for the case $r_+\neq r_-$.
   In the case $r_+=r_-$ we have $\Delta(r)=(r-r_+)^2=(r-M)^2$, hence we can continue the calculation 
   above as follows
   \begin{align*}
      A(x) - A_0\ 
      &=\
      \frac{r-r_+}{r^2+a^2} \left[
	 mr_+ + \frac{ 2a\kappa r_+ + eQ(r_+^2 - a^2) }{r_+^2+a^2}
	 + (r-r_+) \left( m + \frac{ a\kappa + eQr_+ }{r_+^2+a^2} \right)
      \right]
      \\[1ex]
      &=\ \frac{r-r_+}{r^2+a^2} \left[
	 mr_+ -\mu
	 + (r-r_+) \left( m + \frac{ a\kappa + eQr_+ }{r_+^2+a^2} \right)
      \right].
   \end{align*}
      The assertion follows now from \eqref{eq:asymptotics} and \eqref{eq:r asymptotics extreme case}.

      \item
      A simple calculation gives
      \begin{align}
	 \diff{x} A(x)\ &=\ \diff[r]{x}\, \diff{r} A(x(r))\ 
	 =\ \frac{\Delta(r)}{r^2+a^2}\,\diff{r} A(x(r)) \nonumber
	 \\[2ex]
	 & \label{A'(x)}=\ \frac{\sqrt{\Delta(r)}}{r^2+a^2}
	 \left\{
	    m\left(\frac{\Delta(r)}{r^2+a^2} + \frac{r\Delta'(r)}
	       {2(r^2+a^2)} - \frac{2r^2\Delta(r)}{(r^2+a^2)^2} \right)
	    + \frac{2(a\kappa+eQr) r \sqrt{\Delta(r)}}{(r^2+a^2)^2}
	    -\frac{eQ\sqrt{\Delta(r)}}{r^2+a^2} 
	 \right\}
	 \\[2.5ex]
	 & =\ \left\{\begin{alignedat}{3}
	       &O((r-r_+)^{1/2})\ &=\ & O(\exp{[(1/2)\alpha x]}), \hspace{3ex} &(r_+\neq r_-), \\[1ex]
	       &O((r-r+)^2)\ &=\ & O(x^{-2}), \hspace{3ex}  &(r_+ = r_-). 
	    \end{alignedat}\right. \nonumber
	 \end{align}
      as $x \rightarrow -\infty$. 
      \begin{align}
	 \nonumber
	 \diff{x} B(x)\ &=\ \diff[r]{x}\, \diff{r} B(x(r))\ =\
	 \frac{\Delta(r)}{r^2+a^2}\,\diff{r} B(x(r))
	 \\[2ex]
	 \label{eq:BDerivative}
	 &=\ \lambda\, \frac{\sqrt{\Delta(r)}}{r^2+a^2}\, 
	 \left\{
	    \frac{\Delta'(r)}{2(r^2+a^2)}
	    - \frac{2r\Delta(r)}{(r^2+a^2)^2}
	 \right\}
	 \\[2ex]
	 \nonumber
	 &=\
	 \left\{\begin{alignedat}{3}
	       &O((r-r_+)^{1/2})\ &=\ & O(\exp{[(1/2)\alpha x]}), \hspace{3ex} &(r_+\neq r_-), \\
	       &O((r-r+)^2)\ &=\ & O(x^{-2}), \hspace{3ex} &(r_+ = r_-). 
	    \end{alignedat}\right. 
      \end{align}
      as $x \rightarrow -\infty$. 

      \item
      Since
      \[\diff[r]{x}= \frac{\Delta(r)}{r^2+a^2} \sim 1, \ \ x \sim r \ \ 
	 {\rm  as} \ \ x \rightarrow +\infty,\]
      and 
      \[2(r^2+a^2)\Delta(r)+ r(r^2+a^2)\Delta'(r)-4r^2\Delta(r)
	 = 2Mr^3 + 2r^2(a^2-Q^2) - 6 a^2 M r + 2a^2(a^2+Q^2)
	 \sim r^3 
	 \text{ as } r\rightarrow\infty 
	 \]
     the assertion follows from (\ref{A'(x)}).
     The proof of the assertion concerning $B'(x)$ follows directly from~\eqref{eq:BDerivative}.
      \qedhere
   \end{enumerate}
\end{proof}


\section{Absolutely continuous spectrum} 	
\label{sec:AbsContSpec}
  
The following proposition has been shown by \citet{BM}.
\par \bigskip
  
\begin{proposition} $\sigma_{{\rm ess}}(H)={\mathbb R}$.\hfill\qed
\end{proposition}

\par 
We point out that the proof of the proposition relies on the fact that
\begin{equation} \label{asymptotics of V(x)}
   \lim_{x \rightarrow -\infty} [V(x)-A_0I_2] =0,
\end{equation}
where $I_2$ is the  $2 \times 2$ unit matrix.   
Lemma~\ref{lemma:A' for -infty} yields the following theorem.

\begin{theorem}\label{theorem:AbsContSpec}
   \begin{enumerate}
      \item\label{item:AbsContSpecI}
      $H$ has purely absolutely continuous spectrum in 
      ${\mathbb R} \setminus \{A_0\}$.

      \item\label{item:AbsContSpecII}
      $H$ has purely absolutely continuous spectrum in 
      $(-\infty,-m)$ and $(m,+\infty)$,
   \end{enumerate}
   \medskip
   that is, $H$ is absolutely continuous in the complement of $[-m,m] \cap \{A_0\}$.
\end{theorem}

\begin{proof}
   \begin{enumerate}
      \item
      Lemma~\ref{lemma:A' for -infty} gives (\ref{asymptotics of V(x)}) and any component of $V'$ is 
      integrable at $-\infty$. 
      Therefore we can prove the theorem in view of \citet{We}, \citet{SK} and also 
      \citet[Theorem 4.18]{T}.

      \item
      The proof is the same as in~\ref{item:AbsContSpecI} by using Lemma~\ref{lemma:A' for -infty}  \ref{item:A' for infty}.
   \end{enumerate}
\end{proof}

\begin{remark}\label{remark:omega=m}
   The above theorem has already been proven by Schmid by different means (see \citet[Corollary 3.4]{S}); 
   in addition, he has shown that neither $\omega=m$ nor $\omega=-m$ is an eigenvalue of $H$ (\cite[Lemma 3.5]{S}). 
\end{remark}

\begin{theorem}\label{theorem:A0NotEigenvalue} 
If $r_+\neq r_-$, then $A_0$ is not an eigenvalue of $H$.
\end{theorem}

\begin{proof}
   Let us assume that $U={}^t(u_1,u_2) \in \hs{3}$ satisfies 
   \[\left(\begin{array}{cc} 
	    A & B-\diff{x}  \\ B + \diff{x} & C 
   \end{array}\right)
   \left(\begin{array}{c} 
   u_1  \\ u_2 \end{array}\right)= A_0  
   \left(\begin{array}{c} u_1  \\ u_2 
   \end{array}\right),\]
that is,
   \begin{equation}\label{Levinson}
      U'= 
      \left(\begin{array}{c} u'_1  
      \\ u'_2 
      \end{array}\right)
      = \left(\begin{array}{cc} -B & A_0-C \\ A-A_0 & B \end{array}\right) 
        \left(\begin{array}{c} u_1  \\ u_2 \end{array}\right).
   \end{equation}

   As seen in section~\ref{sec:operatorH}, $A(x)-A_0$, $B(x)$ and $C(x)-A_0$ decay exponentially as $x \rightarrow -\infty$. Therefore, 
   the Levinson theorem (e.g. \citet{E}) gives that there are two linearly independent 
   solutions $U^1$, $U^2$ of (\ref{Levinson}) such that
   \begin{eqnarray*}
   & & U^1(x) = \left\{\left(\begin{array}{c} 1  \\ 0 \end{array}\right)
   +o(1)\right\}\ \ {\rm as} \ \ x \rightarrow -\infty,\\
   & & U^2(x) = \left\{\left(\begin{array}{c} 0  \\ 1 \end{array}\right)
   +o(1) \right\}\ \ {\rm as} \ \ x \rightarrow -\infty. 
   \end{eqnarray*}
Hence there are constants $c_1$ and $c_2$  such that 
$U(x)=c_1U^1(x)+c_2U^2(x)$, which is square integrable on $(-\infty,0)$ only if $c_1=c_2=0$. 
\end{proof}

As an important corollary we obtain the following theorem.

\begin{theorem}[Non-existence of time-periodic solutions for $r_+ \neq r_-$]
   \label{theorem:nonexistence}
   Let $\widehat\Psi\in C^1( (r_0,\infty)\times(0,\pi)\times [-\pi,\pi]\times\mathbb R,\, \mathbb C^4)$ 
   be a solution of \eqref{time-dependent equation} satisfying the periodicity conditions
   \begin{align} \label{periodic condition}
      \begin{aligned}
	 \widehat{\Psi}(r,\theta,\varphi,t)
	 &= \widehat{\Psi}(r,\theta,\varphi, t+ \textstyle\frac{2\pi}{\omega_0}), \\ 
	 \widehat{\Psi}(r,\theta,\pi,t) &= \widehat{\Psi}(r,\theta,-\pi, t),
      \end{aligned}
      \hspace{8ex}
      (r,\theta,\varphi,t)\in\widehat\Omega
   \end{align}		
   for some $\omega_0>0$.
   Furthermore, assume that 
   for all $(\varphi,\, t)\in[-\pi,\, \pi]\times\mathbb R$ we have
    $\widehat\Psi(\,\cdot\,,\,\cdot\,,\,\varphi\,,\,t)\in C^0((r_0,\infty)\times(0,\pi);\ \hs{2})$  
    where
   \[\hs{2} := \Ltwo\left((r_+,\infty)\times(0,\pi) ;\ 
	 \textstyle\frac{r^2+a^2}{\Delta(r)} \sin{\theta} \, dr\,d\theta\right)^4 \] 
   with the norm denoted by $\|\cdot\|_{\hs{2}}$ and the inner product denoted by $\langle\cdot,\cdot \rangle_{\hs{2}}$.  

   If $r_-\neq r_+$, then $\widehat \Psi\equiv 0$.
\end{theorem}

\begin{proof}

Let $\widehat{\Psi}$ be a time-periodic solution satisfying the conditions of the theorem.
   Then $\widehat{\Psi}(\,\cdot\,,\,\cdot\,,\,\cdot\,,\,t)$ is an $\hs{1}$-valued  strongly 
   continuous function with respect to $t$ since $\widehat{\Psi}(\,\cdot\,,\,\cdot\,,\varphi,t)$ is 
   uniformly continuous in $\hs{2}$ with respect to $(\varphi,t) \in [-\pi,\pi] \times \mathbb R$.
   Therefore we can expand  $\widehat{\Psi}(r,\theta,\varphi, t)$ as the Fourier series with respect to $t$  
   \[\widehat{\Psi}(r,\theta,\varphi, t)= 
      \sum_{\nu \in {\mathbb Z}} \exp{(-i\omega_0\nu t)} 
      \Psi^{\nu}(r,\theta,\varphi)\]
   strongly in $\Ltwo([0,2\pi/\omega_0];\ \hs{1}\,;\,dt)$, where 
    \[ \Psi^{\nu}(r,\theta,\varphi)= \frac{\omega_0}{2\pi}\int_0^{2\pi/\omega_0} \exp{(i\omega_0\nu t)} 
       \widehat{\Psi}(r,\theta,\varphi,t)\, dt, \qquad\quad
       \sum_{\nu} \|\Psi^\nu \|^2_{\hs{1}}<\infty. \]
   Moreover, each $\Psi^{\nu}(r,\theta,\varphi)$ can be expanded as 
      \[\Psi^{\nu}(r,\theta,\varphi)=
      \sum_{\kappa \in {\mathbb Z}+\frac{1}{2}} 
      \exp{(-i\kappa\varphi)} 
      \Psi^{\nu,\kappa}(r,\theta)\]
   strongly in $\Ltwo([-\pi,\pi]\,;\ \hs{2};\ d\varphi)$ with
   \[ \Psi^{\nu}(\theta,\varphi)= \frac{1}{2\pi}\int_{-\pi}^{\pi} \exp{(i\kappa\varphi)} 
      \Psi^\nu(r,\theta,\varphi)\, d\varphi, \qquad\quad
      \sum_{\kappa} \|\Psi^{\nu,\kappa} \|^2_{\hs{2}}<\infty. \]
      
    For every $\Phi\in\Cinfty((r_0,\infty)\times(0,\pi))^4$ of the form
    \[
       \Phi(r,\theta)\, =\,
       \begin{pmatrix} 
	  \rho_-(r)\, \eta_2(\theta) \\ \rho_+(r)\, \eta_1(\theta) \\
	  \rho_+(r)\, \eta_2(\theta) \\ \rho_-(r)\, \eta_1(\theta) 
           \end{pmatrix} \]
    with $\rho_\pm(r) \in \Cinfty((r_+,\infty))$, $\eta_1(\theta)$, $\eta_2(\theta)$ $\in \Cinfty((0,\pi))$ 
   we obtain, by using (\ref{periodic condition}),
   {\allowdisplaybreaks
      \begin{eqnarray}
	 0 &=& \int_0^{2\pi/\omega_0}dt \int_{-\pi}^\pi d\varphi  
	 \left\langle \textstyle (\widehat{{\cal R}}+\widehat{{\cal A}})\widehat{\Psi},\
	 \exp{(-i\omega_0 \nu t)} \exp{(-i\kappa \varphi)}\Phi \right\rangle_{\hs{2}} \nonumber \\
	 &=& \int_0^{2\pi/\omega_0}dt \int_{-\pi}^\pi d\varphi  
	 \left\langle \widehat{\Psi},\ (\widehat{{\cal R}}+\widehat{{\cal A}})^*
	 \exp{(-i\omega_0 \nu t)} \exp{(-i\kappa \varphi)}\Phi\right\rangle_{\hs{2}} \nonumber \\
	 &=& \int_0^{2\pi/\omega_0}dt \int_{-\pi}^\pi d\varphi  
	 \left\langle \widehat{\Psi},\ \exp{(-i\omega_0 \nu t)} \exp{(-i\kappa \varphi)}
	 ({\cal R}_{\nu,\kappa}+{\cal A}_{\nu,\kappa})^*\Phi\right\rangle_{\hs{2}} \nonumber \\
	 &=& \int_0^{2\pi/\omega_0}  
	 \left\langle \widehat{\Psi},\ \exp{(-i\omega_0 \nu t)} \exp{(-i\kappa \varphi)}
	 ({\cal R}_{\nu,\kappa}+{\cal A}_{\nu,\kappa})^*\Phi\right\rangle_{\hs{1}} \,dt \nonumber \\
	 &=& \frac{2\pi}{\omega_0} \int_{-\pi}^{\pi} \left\langle\Psi^{\nu},\ \exp{(-i\kappa \varphi)}({\cal R}_{\nu,\kappa}^*+{\cal A}_{\nu,\kappa}^*)\Phi
	 \right\rangle_{\hs{2}} \,d\varphi \nonumber \\
	 &=&\frac{4\pi^2}{\omega_0}\left\langle\Psi^{\nu,\kappa},\ ({\cal R}_{\nu,\kappa}^*+ A_{\nu,\kappa}^*)\Phi
	 \right\rangle_{\hs{2}}.  \label{nu and kappa equation}
      \end{eqnarray}   
   }
   In the above calculation, the superscript $\ ^*$ denotes the formal adjoint operator.
   As in~\eqref{angular eigenfunctions}, for fixed $\nu$ and $\kappa$ let
    \[g^{\nu,\kappa,n}(\theta)=
       \left(\begin{array}{c} g_1^{\nu,\kappa,n}(\theta) \\[1ex]
	     g_2^{\nu,\kappa,n}(\theta)  
       \end{array}\right) 
       \hspace{5ex}
       (n \in {\mathbb Z}\setminus \{0\})
    \]
   be a complete family of orthonormal eigenfunctions of $A_{\nu,\kappa}$ with  eigenvalues 
   $\lambda_{\nu,\kappa,n}$, respectively.
   Then $\Psi^{\nu,\kappa}(r,\theta)$ 
   can be expanded in terms of $g^{\nu,\kappa,n}$ $(n \in \mathbb Z\setminus\{0\})$ as follows
   \begin{eqnarray*}
      \Psi^{\nu,\kappa}(r,\theta)
     &=&  \sum_{n \in \mathbb Z\setminus\{0\}} 
     \left( \begin{array}{c} 
	    X^{\nu,\kappa,n}_-(r)\, g_2^{\nu,\kappa,n}(\theta) \\ 
	    X^{\nu,\kappa,n}_+(r)\, g_1^{\nu,\kappa,n}(\theta) \\ 
	    X^{\nu,\kappa,n}_+(r)\, g_2^{\nu,\kappa,n}(\theta) \\ 
	    X^{\nu,\kappa,n}_-(r)\, g_1^{\nu,\kappa,n}(\theta)  
      \end{array}\right),
   \end{eqnarray*}
   where the series converges in the strong sense in $\Ltwo((0,\pi);\ \sin{\theta}\,d\theta))^4$ and   
   $X^{\nu,\kappa,n}_\pm(r)$ satisfies 
   \[\sum_{n \in \mathbb Z\setminus\{0\}}\int_{r_+}^{\infty} \bigl(|X_+^{\nu,\kappa,n}(r)|^2+|X_-^{\nu,\kappa,n}(r)|^2\bigr)\,
   \frac{r^2+a^2}{\Delta(r)}\,dr\, <\, \infty.\]
   Since $\mathfrak A_{\nu,\kappa}$ on $\Cinfty((0,\pi))^2$ is essentially selfadjoint in 
   $\Ltwo((0,\pi);\ \sin{\theta}\,d\theta)^2$, for any $g^{\nu,\kappa,n}(\theta)$ there exists a convergent sequence 
   $(\eta^\ell)_{\ell\in\mathbb N} \subset \Cinfty((0,\pi))^2$ such  that  
   \[A_{\nu,\kappa} \eta^\ell = A_{\nu,\kappa}\left(\begin{array}{c} \eta_1^\ell \\ \eta_2^\ell
                   \end{array}\right)
     \longrightarrow A_{\nu,\kappa} g^{\nu,\kappa,n}=\lambda_{\nu,\kappa,n}g^{\nu,\kappa,n} \]                 
  in $\Ltwo((0,\pi);\ \sin{\theta}\,d\theta)^2$. 
  Substituting $\Phi$ in \eqref{nu and kappa equation} by
  \begin{align*}
     \Phi^\ell(r,\theta)=
     \begin{pmatrix} 
	\rho_-(r)\, \eta_2^\ell(\theta) \\ 
	\rho_+(r)\, \eta_1^\ell(\theta) \\
	\rho_+(r)\, \eta_2^\ell(\theta) \\
	\rho_-(r)\, \eta_1^\ell(\theta) 
     \end{pmatrix} 
  \end{align*}
  and taking the limit $\ell \rightarrow \infty$, we have
  \begin{eqnarray*}  
     \left\langle \sum_{m \in \mathbb Z\setminus\{0\}} 
     \begin{pmatrix}
	X^{\nu,\kappa,m}_-(r)\, g_2^{\nu,\kappa,m}(\theta) \\ 
	X^{\nu,\kappa,m}_+(r)\, g_1^{\nu,\kappa,m}(\theta) \\ 
	X^{\nu,\kappa,m}_+(r)\, g_2^{\nu,\kappa,m}(\theta) \\ 
	X^{\nu,\kappa,m}_-(r)\, g_1^{\nu,\kappa,m}(\theta)  
     \end{pmatrix},\
     (R^*_{\nu,\kappa}-\lambda_{\nu,\kappa,n})
     \begin{pmatrix} 
	\rho_-(r)\, g_2^{\nu,\kappa,n}(\theta) \\ 
	\rho_+(r)\, g_1^{\nu,\kappa,n}(\theta) \\
	\rho_+(r)\, g_2^{\nu,\kappa,n}(\theta) \\ 
	\rho_-(r)\, g_2^{\nu,\kappa,n}(\theta) 
     \end{pmatrix}  \right\rangle_{\hs{2}}
     & = & 0,
  \end{eqnarray*}
   which gives  
   \begin{align*} 
    0\, =\, \int_{r_+}^{\infty}\,\rd r\
     \frac{r^2+a^2}{\Delta(r)}\ \left\langle
   \begin{pmatrix} X_+^{\nu,\kappa,n} \\ X_-^{\nu,\kappa,n} 
   \end{pmatrix} , 
   \begin{pmatrix}  imr-\lambda_{\nu,\kappa, n} & {\cal D}_{+,\nu, \kappa}^*\sqrt{\Delta}\\
   {\cal D}_{-,\nu, \kappa}^*\sqrt{\Delta} & -imr-\lambda_{\nu,\kappa, n}
   \end{pmatrix} 
   \begin{pmatrix} \rho_+ \\ \rho_-  
   \end{pmatrix} \right\rangle_{\mathbb C^2},
   \end{align*}
   which implies
   \[
   \left(\begin{array}{cc} 
	 -imr-\lambda_{\nu,\kappa, n} & \sqrt{\Delta}\,{\cal D}_{-,\nu, \kappa} \\[1ex]
	 \sqrt{\Delta}\,{\cal D}_{+,\nu, \kappa} & imr -\lambda_{\nu,\kappa, n}  
   \end{array}  \right)
   \left(\begin{array}{c} 
	 X^{\nu,\kappa,n}_+(r) \\[1ex] X^{\nu,\kappa,n}_-(r)   
   \end{array} \right)
   = 0.
   \]   
   If we set (cf. \eqref{eq:widetilde f} and \eqref{f int. cond})
   \[
      f^{\nu,\kappa,n}(x) = 
      \left(\begin{array}{c} 
	    f^{\nu,\kappa,n}_1(r(x)) \\ f^{\nu,\kappa,n}_2(r(x))   
      \end{array}\right)
      = \frac{1}{\sqrt{2}} \left(\begin{array}{cc}
	    -\i & \i  \\ -1 & -1    
      \end{array}\right)
      \left( \begin{array}{c} 
	    X^{\nu,\kappa,n}_+(r(x)) \\ X^{\nu,\kappa,n}_-(r(x))
      \end{array}\right),
      \] 
   we have
   \[ H_{\nu,\kappa,n} f^{\nu,\kappa,n}=\omega_0 \nu  f^{\nu,\kappa,n}\] 
   (see (\ref{definition of H})).
   Then, Theorem~\ref{theorem:AbsContSpec} and Theorem~\ref{theorem:A0NotEigenvalue} give $ f^{\nu,\kappa,n}=0$ 
   for any $n \in \mathbb Z\setminus\{0\}$ 
   and $\kappa \in {\mathbb Z}+(1/2)$, which yields $\Psi^{\nu,\kappa}(r,\theta)=0$.   
\end{proof}

The non-existence of time-periodic solutions is shown by \citet{FKSY} by different means.


\section{The case $r_+=r_-$}	
\label{sec:ExtremeCase}
In the previous section we have seen that there are no eigenvalues of \eqref{time-dependent equation} 
in the case $r_+\neq r_-$.
In this section we discuss whether $A_0$ is an eigenvalue of $H$ in the case $r_+=r_-$.
Recall that in this case  the function $\Delta$ has only one zero and that $r_+ = r_- = M$. 
\par \bigskip

\begin{theorem}\label{theorem:EigenvalueCondition}
   If 
   \[\lambda^2+m^2M^2 - \mu^2 \leq \frac{\,1\,}{4}, \] 
   then $A_0$ is not an eigenvalue of $H$.
\end{theorem}

\begin{proof}
   Let us assume that $U={}^t(u_1,u_2) \in \hs{3}$ satisfies 
(\ref{Levinson}),
that is,
   \[
      U'= 
      \left(\begin{array}{c} u'_1  
      \\ u'_2 
      \end{array}\right)
      = \left(\begin{array}{cc} -B & A_0-C \\ A-A_0 & B \end{array}\right) 
        \left(\begin{array}{c} u_1  \\ u_2 \end{array}\right).
   \]
Lemma~\ref{lemma:A' for -infty} yields for $x\rightarrow -\infty$
   \[U'(x)
    = -\frac{1}{x} 
      \left(\begin{array}{cc} \lambda & mM+\mu \\ 
                              mM-\mu & -\lambda 
      \end{array}\right) U(x) 
      + O\left(x^{-2}\right) U(x).
   \]
If we introduce $s=\log{(-x)}$, we have
\begin{align}\label{DE in s}
   \frac{\rd}{\rd s}\, U(x(s)) =   
   \left(\begin{array}{cc} \lambda & mM+\mu \\ 
	 mM-\mu & -\lambda 
   \end{array}\right) U(x(s))
   + O\left(\exp{(-s)}\right) U(x(s)) \qquad {\rm as}\quad s \rightarrow +\infty.
\end{align}

The matrix 
   \[S := \left(\begin{array}{cc} \lambda & mM+\mu \\ 
     mM-\mu & -\lambda \end{array}\right) 
   \]
can  be diagonalised by a non-singular matrix $T$ as
   \[T^{-1}ST=  
      \left(\begin{array}{cc} \sqrt{\lambda^2+m^2M^2-\mu^2} & 0 \\
       0 & -  \sqrt{\lambda^2+m^2M^2-\mu^2} \end{array}\right)
   \]
if $\lambda^2+m^2M^2-\mu^2\neq 0$. If $\lambda^2+m^2M^2-\mu^2= 0$, its Jordan canonical form is
 \[T^{-1}ST=  
      \left(\begin{array}{cc} 0 & 1 \\
       0 & 0 \end{array}\right).     
 \]

  According to Theorem 1.8.1 and Theorem 1.10.1 in \citet{E},
we have two linearly independent solutions $U^1(s)$, $U^2(s)$ of \eqref{DE in s} such that
if $\lambda^2+m^2M^2-\mu^2\neq 0$,
   \begin{alignat*}{3}
   U^1(s)\ &=\ \{v_1+o(1)\} \exp{(\sqrt{\lambda^2+m^2M^2-\mu^2}\,s)} \
          &&=\ \{v_1+o(1)\} (-x)^{\sqrt{\lambda^2+m^2M^2-\mu^2}}, \\[1ex]
   U^2(s)\ &=\ \{v_2+o(1)\} \exp{(-\sqrt{\lambda^2+m^2M^2-\mu^2}\,s)} \
          &&=\ \{v_2+o(1)\} (-x)^{-\sqrt{\lambda^2+m^2M^2-\mu^2}} 
   \end{alignat*}
   as $x \rightarrow -\infty$ and, if $\lambda^2+m^2M^2-\mu^2 = 0$,   
    \begin{alignat*}{3}
   U^1(s)\ &=\  v_1+o(1),   \
             \\[1ex]
   U^2(s)\ &=\ sv_1 + v_2 + o(s)  \
            =\ v_1\log{(-x)}+v_2 + o(\log{(-x)}) 
   \end{alignat*}
   as $x \rightarrow -\infty$, where $T=(v_1,v_2)$. 

   Therefore a necessary condition for the existence of an $\Ltwo(-\infty,0)$-eigenfunction is
   \[\sqrt{\lambda^2+m^2M^2-\mu^2}\, \in\, (1/2,\,\infty), 
      \qquad {\rm i.e.}\quad
      \lambda^2+m^2M^2-\mu^2\, >\, 1/4.
      \qedhere  \]
\end{proof}

\begin{remark}
The above discussions show that for $\omega$ to be an 
eigenvalue of $H$ the following conditions are necessary:
   \begin{eqnarray}
      r_+ = r_- = M      &&   
      ({\rm Theorem~\ref{theorem:A0NotEigenvalue}}),  \label{4.3.1}  \\
      \omega = A_0 = -\frac{a\kappa + eQr_+}{r_+^2+a^2}       &&   
      ({\rm Theorem~\ref{theorem:AbsContSpec}\,\ref{item:AbsContSpecI}}),  \label{4.3.2}  \\ 
      \omega^2 < m^2  &&   
      ({\rm Theorem~\ref{theorem:AbsContSpec}\,\ref{item:AbsContSpecII}),\ \ Remark~\ref{remark:omega=m}}),  
      \label{4.3.3}  \\
      \lambda^2+m^2M^2-\mu^2 > \frac{1}{4} &&   
      ({\rm Theorem~\ref{theorem:EigenvalueCondition}}).  \label{4.3.4}
   \end{eqnarray} 
\end{remark}
However, the solvability of the above system is not yet sufficient for the existence of an energy eigenvalue.

\citet{S} showed that if in addition either\\
\hspace*{\fill}or
\parbox[c]{.9\textwidth}{%
   \begin{align}
      \refstepcounter{equation}
      \tag{\theequation.0} \label{sufficient 0}
      &\beta-\sigma\lambda=0, \hspace{5ex} \alpha+\eta=0 \\[2ex]
      \tag{\theequation.N} \label{sufficient N}
      &N+\alpha+\eta=0
      \qquad \text{for some positive integer } N,
   \end{align}
}
\\
holds, then the solvability  of the system \eqref{4.3.2}--\eqref{sufficient N} is sufficient for the existence of an eigenvalue $\omega$ of $H$.

In the above formulae, we used
\begin{gather*}
   \sigma :={\rm sign}\,\omega, \hspace{4ex}
   \eta    :=\sqrt{\lambda^2 + M^2m^2 - \mu^2}, \hspace{4ex}
   \mu    = 2M\omega + eQ = -\frac{2a\kappa M}{M^2+a^2} - eQ\,\frac{M^2-a^2}{M^2+a^2}, 
   \\[2ex]
   \alpha :=\frac{M m^2-\omega\mu}{\sqrt{m^2-\omega^2}}, \hspace{4ex}
   \beta  :=\frac{(M|\omega|-\sigma\mu)m}{\sqrt{m^2-\omega^2}}.
\end{gather*}
Note that the variable $\eta$ is denoted by $\kappa$ in \citep{S}.


\section{Energy eigenvalues in the case $r_+=r_-$}	
\label{sec:energy_eigenvalues}
If $\omega$ is an energy eigenvalue of~\eqref{time-dependent equation}, then there must be 
an $N\in\mathbb N_0$ such that $\omega$ satisfies the complicated system of conditions 
\eqref{4.3.2}--\eqref{4.3.4} and \eqref{sufficient N}.
It is not clear that for given data of the  black hole and the particle there are tuples
$(\nu\omega_0,\, \kappa,\, \lambda_{\nu,\kappa,n})$ that solve the system~\eqref{4.3.2}--\eqref{sufficient N}.
\citet{S} has shown that in the so-called Kerr case (i.e. if $Q=0$) for fixed data 
of the spin-$\frac{1}{2}$ particle there exist two sequences $(a^\pm_N)_{N\in\mathbb N}$ 
such that for $a=a_N^\pm$ the value $\omega_N = -\frac{\eta}{2a_N^\pm}$ is an energy eigenvalue. 

\medskip
Here, we fix the black hole data $M,\ Q$ and $a$ and vary the mass of the fermion to obtain 
the existence of energy eigenvalues in the case $r_-=r_+$.

\begin{theorem}
   \label{theorem:existence}
   Fix $M>0$, $a,\ Q,\ e \in\mathbb R$, $\nu \in {\mathbb Z}$ and $\kappa\in\mathbb Z+\frac{1}{2}$.
   Let $\omega:= -\frac{\kappa a + eQM}{a^2+M^2}$. 
   Take $\lambda=\lambda_{\nu,\kappa,n}$ for any sufficiently large $|n|.$   
   If $\omega(eQ+M\omega) \le 0$ then there are no energy eigenvalues of $H_{\nu,\kappa,n}$.
   If $\omega(eQ+M\omega) > 0$ then there is a sequence $(m_N)_{N\in\mathbb N}\subseteq(|\omega|,\,\infty)$ 
   such that if $m\in\{m_N\,:\, N\in\mathbb N\}$, then 
   $\omega$ is an energy eigenvalue of $H_{\nu,\kappa,n}$.
For $N_0$ large enough, the sequence $(m_N)_{N\ge N_0}$ is monotonously decreasing and converges to $|\omega|$.
\end{theorem}

Before we prove the theorem, let us emphasise that $\lambda=\lambda_{\nu,\kappa,n}$ does depend also on $m$. 
Therefore, we denote it by $\lambda=\lambda_{\nu,\kappa,n}(m)$.
In order to check the condition (\ref{4.3.4}) we prepare the following lemma. 

\begin{lemma} 
   \label{lemma:check_condition}
   Fix $\nu\in\mathbb Z$ and $\kappa\in \mathbb Z + \frac{1}{2}$. 
   If $|n|$ is sufficiently large, then the inequality $\eqref{4.3.4}$ holds for any $m>0$, 
   $M>0$ and $\mu \in {\mathbb R}$, that is,
   \[\lambda_{\nu,\kappa,n}^2+m^2M^2-\mu^2\ >\ \frac{1}{4}. \]
\end{lemma} 
\begin{proof}

   It follows from standard perturbation theory (applied to the angular operator with $m$ as perturbation parameter, see \citet{W}, \citet{K}) that 
   \begin{align*}
      \left|\frac{\rd}{\rd m} \lambda_{\nu,\kappa,n}(m) \right|
      \ \le\
      \left\| \begin{pmatrix} -a\cos\theta & 0 \\ 0 & a\cos\theta \end{pmatrix}
      \right\|
      \ =\ |a|,
   \end{align*}
   hence, for $m>|\omega|$,  
   \begin{equation} \label{perturbation}
      | \lambda_{\nu,\kappa,n}(|\omega|) | - |a|(m-|\omega|)
      \ \le\
      |\lambda_{\nu,\kappa,n}(m)| 
      \ \le\
      | \lambda_{\nu,\kappa,n}(|\omega|) | + |a|(m-|\omega|).
   \end{equation}
   \newcommand{\wm}{\widetilde m}
   Let $\wm = M^{-1}\sqrt{\mu^2 + 1/4}$.
   Since the sequence $(\lambda_{\nu,\kappa,n})_{n}$ is monotonously increasing and unbounded from below and from above, there is an integer $n_0$ such that 
   \begin{align*}
      | \lambda_{\nu,\kappa,n}(|\omega|) |\ >\ 
      |a|(\wm-|\omega|) + \sqrt{\mu^2 + 1/4}
   \end{align*}
   for all $|n| \geq n_0$. If $m\in\bigl(|\omega|,\, \wm\bigr]$, we have
   \begin{eqnarray*}
   |\lambda_{\nu,\kappa,n}(m)| &\geq&  | \lambda_{\nu,\kappa,n}(|\omega|) | - |a|(m-|\omega|)\\
   &\geq&  | \lambda_{\nu,\kappa,n}(|\omega|) | - |a|(\wm-|\omega|)> \sqrt{\mu^2+1/4}
   \end{eqnarray*}
   which implies 
   \begin{align*}
      \sqrt{\lambda_{\nu,\kappa,n}(m)^2 + M^2m^2 - \mu^2}\ \geq\ \sqrt{\mu^2+1/4+ m^2M^2 -\mu^2}> 1/2.
   \end{align*}
   If $m > \wm = M^{-1}\sqrt{\mu^2 + 1/4}$, then we have
   \begin{align*}
      \sqrt{\lambda_{\nu,\kappa,n}(m)^2 + M^2m^2 - \mu^2} 
      \ &>\
      \sqrt{M^2\wm^2 - \mu^2} 
      \ =\ 1/2.
      \qedhere
   \end{align*}
\end{proof}
Now we shall prove Theorem~\ref{theorem:existence}.
\begin{proof}[Proof of Theorem~\ref{theorem:existence}]
   By definition, $\omega$ satisfies condition~\eqref{4.3.2}.
   As seen in Lemma~\ref{lemma:check_condition}, there is an $n_0\in\mathbb N$ such that condition ~\eqref{4.3.4} is satisfied for all $|n|\geq n_0$.
   From now on, let us assume that condition~\eqref{4.3.4} holds.
   Next we consider the conditions~\eqref{sufficient 0} and \eqref{sufficient N}.
   To this end we compute
   \begin{align*}
      \alpha + \eta\ &=\ 
      \frac{M m^2 - \omega\mu}{\sqrt{m^2-\omega^2}} 
      + \sqrt{\lambda_{\kappa,n}^2 + M^2m^2 - \mu^2}\\
      &=\ 
      -\frac{\omega(eQ+M\omega)}{\sqrt{m^2-\omega^2}} + M\sqrt{m^2-\omega^2}
      + \sqrt{\lambda_{\kappa,n}^2 + M^2m^2 - \mu^2}.
   \end{align*}
   If $\omega(eQ+M\omega)\le 0$, then $\alpha + \eta > \frac{1}{2}$ and condition \eqref{sufficient N} 
   cannot be satisfied for any $N\in\mathbb N_0$.
   Assume now that $\omega(eQ+M\omega)>0$. Then the function 
   \begin{eqnarray*}
    A: (|\omega|,\, \infty) &\longrightarrow& \mathbb R,\hspace{2ex} \\
    m &\mapsto&
      -\frac{\omega(eQ+M\omega)}{\sqrt{m^2-\omega^2}} + M\sqrt{m^2-\omega^2}
      + \sqrt{\lambda_{\nu,\kappa,n}(m)^2 + M^2m^2 - \mu^2}
   \end{eqnarray*}
   is continuous, satisfies 
   $\lim_{m\searrow |\omega|} A(m) = -\infty$, 
   $\lim_{m\rightarrow \infty} A(m) = \infty$
   in view of (\ref{perturbation}).
   Hence for every $N\in\mathbb N$ there is at least one $m_N\in(|\omega|, \infty)$ such that 
   $A(m_N) =  -N$ and therefore satisfies condition \eqref{sufficient N}.
   Since the function $A$ is monotonously increasing in an interval $(|\omega|,|\omega|+\delta)$ for a sufficiently small 
   $\delta>0$, 
   it follows that for $N$ large enough there is only one $m_N$ satisfying $A(m_N)= -N$ and that the sequence 
   $(m_N)_N$ is decreasing.
\end{proof}

\begin{remark}
   The proof shows that for fixed $m$ only a finite number of $\lambda_{\nu,\kappa,n}$ is allowed in order to satisfy 
   condition \eqref{sufficient N}.
   The closer $m$ is to $|\omega|$, the more (and the larger) values for $\lambda_{\nu,\kappa,n}$ are allowed.
\end{remark}

\begin{remark}
   The condition $\omega(eQ+M\omega)>0$ is satisfied if the ratio $eQ/\kappa$ is sufficiently small since
   \begin{align*}
      \omega(eQ+M\omega)\ 
	 &=\ -\frac{a}{(a^2+M^2)^2}(\kappa a +eQM)(eQa-\kappa M)\\[1ex]
	 &=\ \frac{a}{(a^2+M^2)^2}[aM\kappa^2 +eQ(M^2-a^2)\kappa -aMe^2Q^2] \\[1ex]
	 &=\ \frac{a^2M}{(a^2+M^2)^2}
	 \left[
	    \left(\kappa - \frac{eQa}{2M} +\frac{eQM}{2a}\right)^2
	    -\frac{e^2Q^2}{4}\left( \left(\frac{a}{M}-\frac{M}{a}\right)^2 + 4\right) 
	 \right] \\[1ex]
	 &=\ - \frac{a^2\kappa^2 M}{4(a^2+M^2)^2}
	 \left[
	    \left(\frac{2eQ}{\kappa} + \left( \frac{a}{M} - \frac{M}{a} \right)\right)^2
	    -\left( 4 + \left(\frac{a}{M}-\frac{M}{a}\right)^2 \right) 
	 \right].
   \end{align*}

\end{remark}

\begin{remark}
   Let $\omega$ be an energy eigenvalue, $m\in \{m_N\, :\, N\in\mathbb N\}$ (see Theorem~\ref{theorem:existence}),
   and $f^{\nu,\kappa,n}$  the eigenfunctions of $H_{\nu,\kappa,n}$. 
   If we set
   \[\left(\begin{array}{c} X_+^{\nu,\kappa,n}(r) \\ X_-^{\nu,\kappa,n}(r) 
      \end{array}\right)
      = \frac{1}{\sqrt{2}} \left(\begin{array}{cc} i & -1  \\ -i & -1 
	    \end{array}\right) \left(\begin{array}{c} f_1^{\nu,\kappa,n}(x) \\ f_2^{\nu,\kappa,n}(x) 
   \end{array}\right)\]
   then 
   \begin{align*}
      \widehat\Psi(r,\theta,\phi,t)\ =\  
      \exp{(-i\omega t)}\,\exp{(-i\kappa\varphi)} 
      \begin{pmatrix} 
	 X_-^{\nu,\kappa,n}(r)\, g_2^{\nu,\kappa,n}(\theta) \\ 
	 X_+^{\nu,\kappa,n}(r)\, g_1^{\nu,\kappa,n}(\theta) \\ 
	 X_+^{\nu,\kappa,n}(r)\, g_2^{\nu,\kappa,n}(\theta) \\ 
	 X_-^{\nu,\kappa,n}(r)\, g_1^{\nu,\kappa,n}(\theta)
      \end{pmatrix}
   \end{align*}
   is a time-periodic solution of~\eqref{time-dependent equation}.

\end{remark}


\begin{acknowledgments}
The authors express their gratitude to the DFG (Deutsche Forschungsgemeinschaft) for giving us the opportunity for our discussions in Japan on the occasion of the German Year in Japan 2005.    
We thank Professor Hubert Kalf and the referee for valuable suggestions and advices.
This work is partially supported by the Grant-in-Aid for Scientific Research (C) No. 18540196, Japan Society for the Promotion of Science.

\end{acknowledgments}


\end{document}